\documentclass[12pt]{article}
\usepackage[dvips]{graphicx}  
\usepackage{epsf}  
\begin{document}
\begin{center}
{\large {\bf Persistence in extended dynamical systems}}
\end{center}
\bigskip
\begin{center} Purusattam Ray$^+$ \end{center}
\vspace {0.2 cm}
\begin{center} 
{\it The Institute of Mathematical Sciences, Chennai 600 113, India}
\end{center}

\bigskip
\bigskip
\bigskip

Persistence in spatially extended
dynamical systems (like coarsening systems and other nonequilibrium 
systems) is reviewed. We discuss, in particular, the spatial 
correlations in the persistent regions and their evolution in 
time in these systems. 
We discuss the dependence of the persistence behavior on 
the dynamics of the system and consider the specific example of 
different updating rules in the temporal evolution
of the system. Lastly, we discuss the universal behavior shown by 
persistence in various stochastic models belonging to the 
directed percolation universality class.

\bigskip


\bigskip


\leftline {$^+$E-mail: ray@imsc.res.in}
\newpage

Persistence has been studied in the past 
in the context of first passage problems in stochastic
processes \cite{chandra}. It tells us how long a stochastic 
variable $X(t)$ retains a property   
as it evolves in time. Persistence is, in general, characterized   
by the persistence probability $P(t)$, defined 
as the probability that $X(t)$ evolving in time retains a particular 
property till time $t$. The property can, for example, be the sign 
of $X(t)$ or crossing the origin etc. 
Persistence tells us how a system
retains its memory as it evolves in time. The study of
persistence is difficult as it involves, in general, the 
knowledge of the entire time evolution of the system.

Analytical treatment of persistence is viable in 
stochastic processes which are Gaussian and stationary. A Gaussian 
process is completely characterized by the two-point 
correlator $f(t,t') = <X(x,t)X(x,t')>$ and for 
a stationary process $f(t,t')$ is a function of $(t-t')$ only. 
Further, if the process is Markovian (in a Markovian process, the 
state $X(t)$ of the variable at time $t$ depends only on its  
state $X(t-1)$ at the preceding time step) the two-point correlator 
takes a simple exponential form and the persistence probability 
is obtained exactly \cite{satya}. In processes which are not  
stationary, the correlator is a function of both $t$ and $t'$. 
Such a process, where the correlator has time 
dependence of the specific form $t/t'$, can be mapped on to a 
corresponding stationary process by a "log-time" transformation.
In the problem of Brownian walker (Gaussian and Markovian but
not stationary), the persistence probability $P(t)$ is known  
exactly: $P(t) \sim t^{-1/2}$ \cite{feller}. 
In non-Markovian processes, the two-point correlator takes 
non-trivial form. The persistence probability $P(t)$ depends 
on the full functional form of $f(t,t')$ and not 
just on its asymptotic form for large $(t-t')$ (this reflects the 
non-Markovian aspect of the process). 
As a result, a closed form expression for $P(t)$  
becomes difficult to obtain.

Recent work on persistence has shown how remarkably  
an interacting nonequilibrium many-body system,
evolving in time, retains memory of its initial state
(see \cite{satya,puru} for a review).
In such systems, persistence is the probability that 
nonequilibrium field $\phi(r,t)$, fluctuating in space
$r$ and time $t$ according to some dynamics, retains some property 
(such as its sign)
for a certain period of time. For example, $\phi$ may
be the coarsening spin field in the Ising model, quenched from
a high temperature to a low temperature. The persistence
probability $P(t)$, in this case, is defined
as the probability that a spin does not flip
till time $t$ \cite{derrida}.
Another example is the diffusing field $\phi(r,t)$ with a
random starting configuration. $P(t)$ is defined here as
the probability that $\phi(r,t)$ remains above or below
the average value of $\phi$ till time $t$ \cite{diff}.
Examples of systems where persistence has been studied
theoretically include fluctuating interfaces \cite{inter},
automaton models and population dynamics \cite{auto} and
reaction-diffusion systems in pure \cite{cardy} and
disordered environments \cite{fisher}. Persistence finds
application in wide varieties of systems from
granular \cite{gran}, chemical \cite{raychaudhuri}
and biological systems \cite{bennaim}, ecology \cite{keeling}
to seismology \cite{lee}. Persistence has been measured in
various experimental systems including breath figures
\cite{marcos}, soap bubbles \cite{tam}, laser polarised
Xe-gas \cite{wong} and liquid crystals \cite{yurke},

In all these cases and in a large class of systems, one
observes a power-law decay of the persistence 
probability $P(t) \sim t^{-\theta}$. The
exponent $\theta$ is called the persistence exponent. 
There have been many attempts in recent years to determine 
the exponent $\theta$ analytically for various systems 
and processes \cite{satya}. Persistence exponents belong to a new class 
of exponents, as it cannot be derived, in general, from 
other static and dynamic exponents. Persistence probes the 
full, in general non-Markovian time evolution of 
a local fluctuating variable, such as a spin or density 
field, from its initial state. Knowing the asymptotic 
properties of the evolution kernel for the time evolution is 
insufficient to evaluate the persistence exponent. In many 
cases progress has been made through controlled expansions 
about Markov processes \cite{satya}. Exact expression for 
$\theta$ is known only in one-dimensional Potts model for 
any Potts state $q$ \cite{derrida2}:

\begin{equation}
\theta(q) = -\frac{1}{8} + \frac{2}{\pi^2}[cos^{-1}(\frac{2-q}{\sqrt{2}q})]^2.
\label{potts}
\end{equation}  

\noindent This gives $\theta(2) = \frac{3}{8}$ as the persistence
exponent for one-dimensional Ising model. In this review, we 
emphasize more on some recent observations on persistence, 
namely: the spatial correlation and dynamical scaling in 
persistence, the dependence of persistence on updating rules
and possible universal behavior of persistence in directed 
percolation problems.

\begin{figure}[tbp]
\vspace*{-0.5in}
\centerline{
\hbox  {
        \vspace*{0.5cm}
        \epsfxsize=6.0cm
        \epsfbox{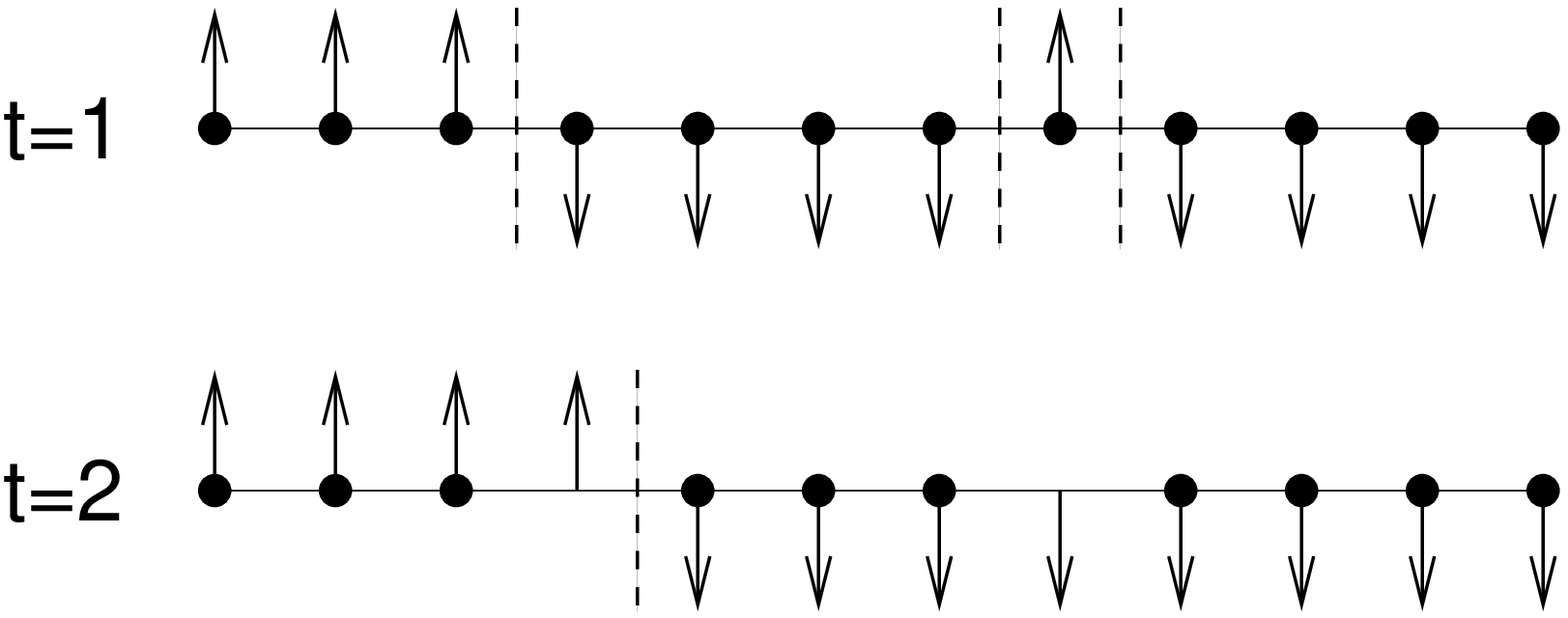}
        \hspace*{1.2cm}
        \epsfxsize=6.0cm
        \epsfbox{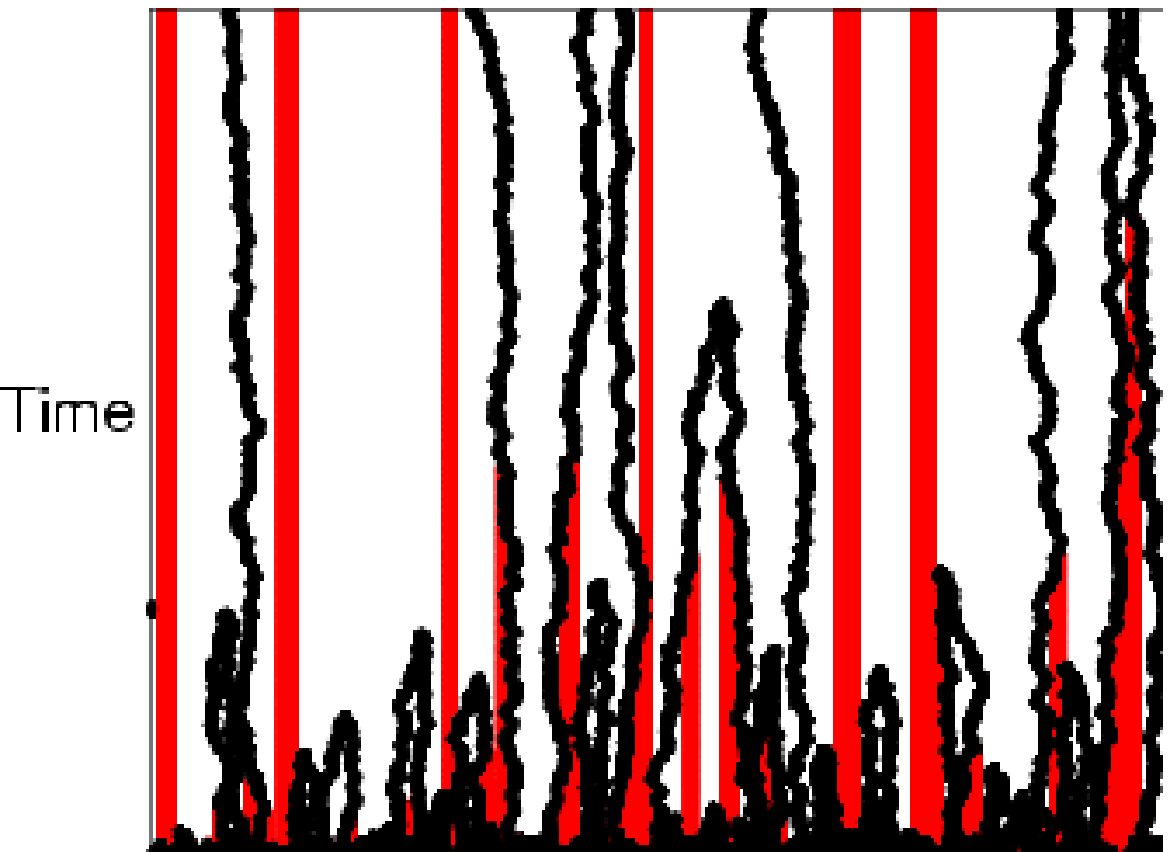}
        \vspace*{0.5cm}
       }
          }

\centerline{
\hbox  {
        \vspace*{0.5cm}
        \epsfxsize=0.5cm
        \epsfbox{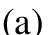}
        \hspace*{7.5cm}
        \epsfxsize=0.75cm
        \epsfbox{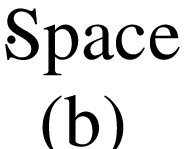}
        \vspace*{0.5cm}
       }
          }

\caption{ (a) Two successive time steps in the evolution of a 
Glauber-Ising chain are shown. Domain
walls are shown as dotted vertical lines and the spins with black 
filled circles at 
their bases are the persistent spins. A domain wall motion 
is associated with a spin 
flip and its loss of persistence. At right two domain 
walls annihilate and two domains 
coalesce. 1(b) World lines of the random walkers and the 
persistent spins. The 
wiggly lines are the domain walls performing random
 walk motion and the vertical 
grey columns are the persistent spins.  }
\label{space}
\end{figure}

In interacting many body systems, the time evolution of 
the field $\phi(x_1)$ at point $x_1$ 
depends on the evolution of the field $\phi(x)$ at a 
different point $x$. A strong correlation 
may arise, as a result, between the persistence of $\phi$ 
at space point $x$ to that at some other point, say, $x+a$.
The persistence properties depend on this correlation as we 
will see in the following. To see how the time evolutions of 
the spin field at two space points can be correlated, consider 
a coarsening Glauber Ising-chain at 
two successive time steps when it is quenched from a high 
temperature to zero temperature (shown in fig.~\ref{space}a). Here, 
the dynamics is as follows: each spin at time $(t+1)$ 
assumes the state of one of its neighboring spins at time 
$t$ with equal probability. It is easy to see that  
a spin flips only when it is crossed by a domain wall.
A spin situated deep within a domain, persists as
long as its neighboring spins persist. This gives
rise to non-trivial correlations in the positions of
the persistent sites \cite{damien} .

Fig.~\ref{space}b displays the time evolution of domain
walls and persistent spins in a Glauber Ising chain quenched to zero
temperature from a random initial spin configuration. The
domain wall is taken to be the mid-point of a bond separating
up and down spin domains. The domain walls perform random walks 
and when two walls meet they annihilate each other and the
system coarsens. As a result, the number of the domain walls
decays with time. Equivalently the average size of domains
increases with time (the average domain size
$L \sim t^{1/2}$ \cite{bray}). A spin remains persistent as long
as it does not encounter a domain wall. The number of persistent
sites decays with time as $t^{-\frac{3}{8}}$. As we will
discuss, the persistent spins at any time slice are strongly
correlated in space and this correlation evolves with time
in accordance with a dynamical scaling law.

\begin{figure}[tbp]
\vspace*{-0.5in}
\centerline{
\hbox  {
        \vspace*{0.5cm}
        \epsfxsize=6.0cm
        \epsfbox{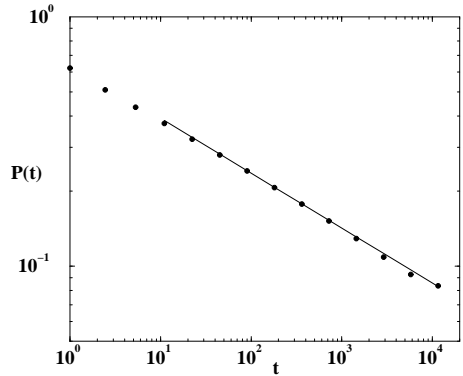}
        \vspace*{0.5cm}
       }
          }
\caption{ The persistence probability $P(t)$ vs. $t$ is shown in 
a logarithmic plot for a $500 \times 500$ Ising model quenched 
from a random initial spin configuration to zero temperature. The 
solid line has a slope 0.22 and is a guide to the eye.}{}
\label{pers}
\end{figure}

\begin{figure}[tbp]
\centerline{
\hbox  {
        \vspace*{0.5cm}
        \epsfxsize=6.0cm
        \epsfbox{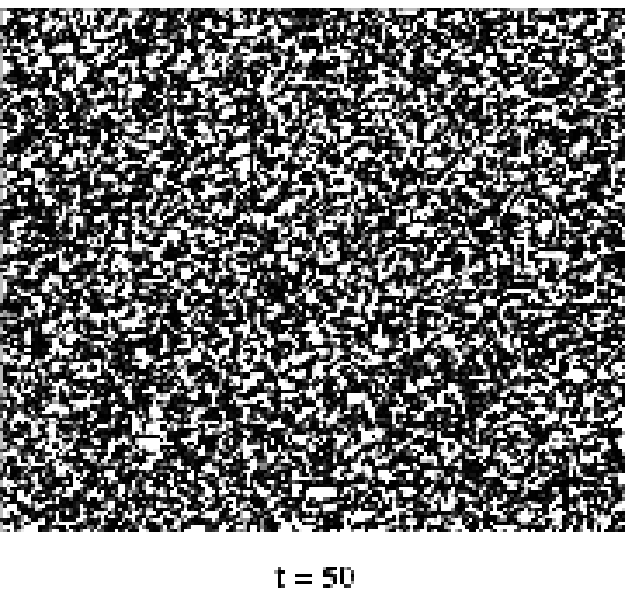}
        \hspace*{1.2cm}
        \epsfxsize=6.0cm
        \epsfbox{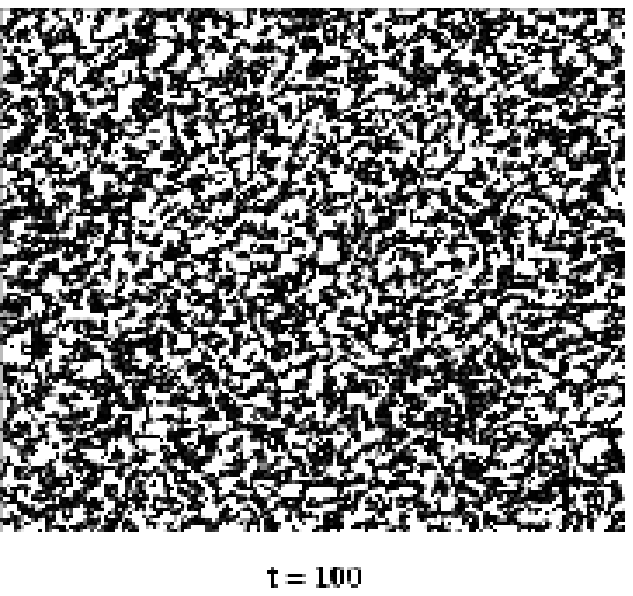}
        \vspace*{0.5cm}
       }
          }

\centerline{
\hbox  {
        \vspace*{0.5cm}
        \epsfxsize=6.0cm
        \epsfbox{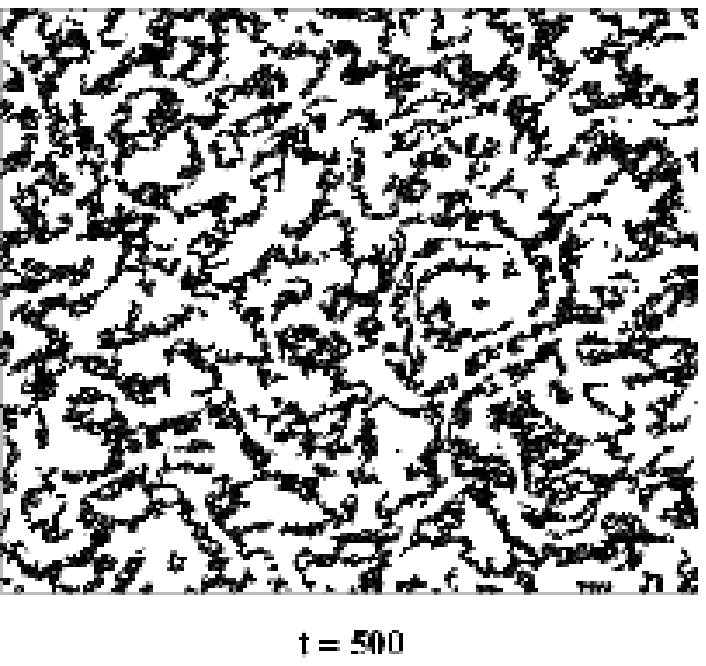}
        \hspace*{1.2cm}
        \epsfxsize=6.0cm
        \epsfbox{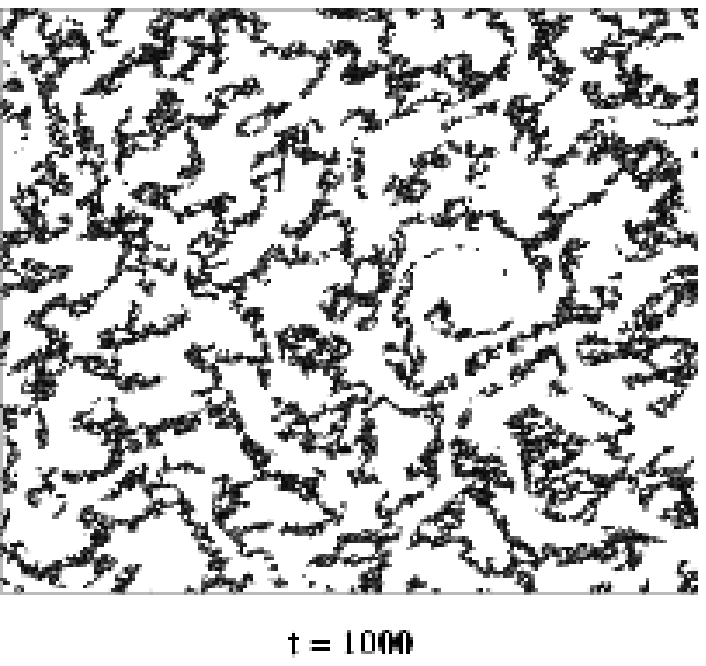}
        \vspace*{0.5cm}
       }
          }
\caption{ Persistent spins in a $500 \times 500$ Ising model at 
Monte Carlo times $t = 50, 100, 500$ and 1000, after the system 
is quenched from an initial random 
spin configuration to zero temperature. }{}
\label{phased}
\end{figure}

\begin{figure}[tbp]
\centerline{
\hbox  {
        \vspace*{0.5cm}
        \epsfxsize=6.0cm
        \epsfbox{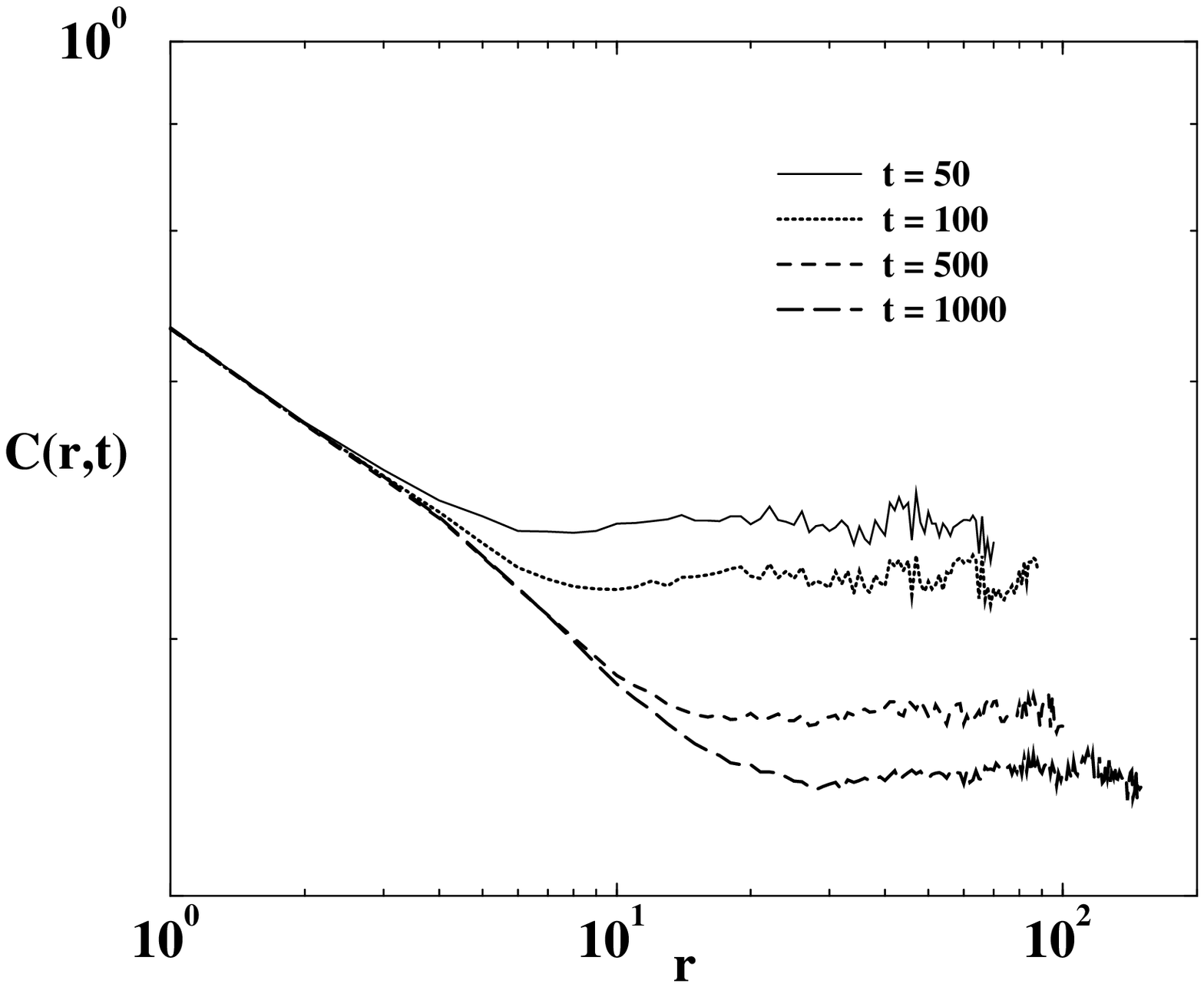}
        \hspace*{1.2cm}
        \epsfxsize=6.0cm
        \epsfbox{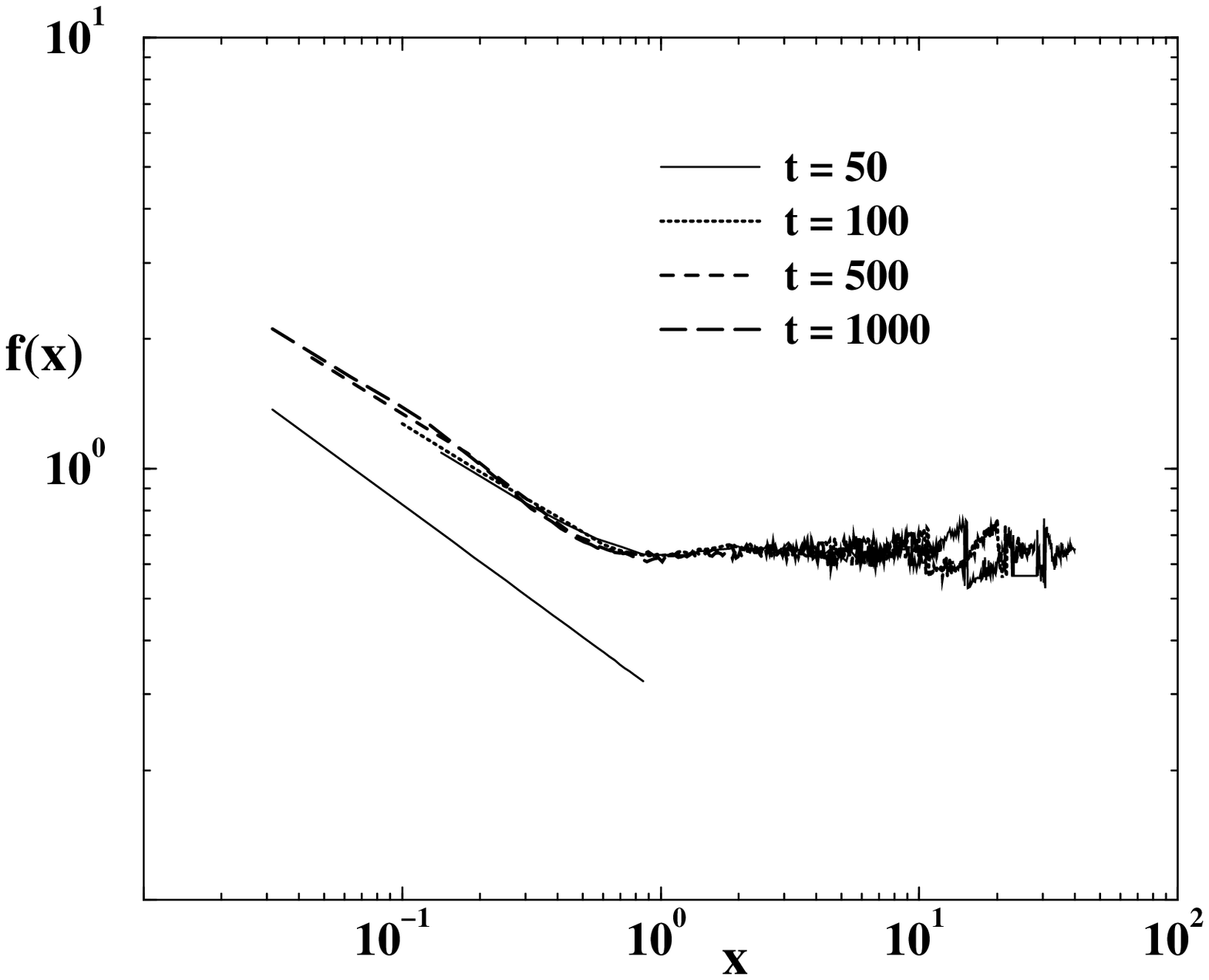}
        \vspace*{0.5cm}
       }
          }  

\centerline{
\hbox  {
        \vspace*{0.5cm}
        \hspace*{1.0cm}  
        \epsfxsize=0.5cm
        \epsfbox{fig1a.eps}
        \hspace*{6.8cm}
        \epsfxsize=0.5cm
        \epsfbox{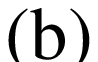}
        \vspace*{0.5cm}
       }
          }

\caption{(a) The two point correlator $C(r,t)$ is plotted 
against $r$ in a logarithmic plot for Monte Carlo times 
$t = 50, 100, 500$ and 1000 (successively from above). 
The system is Ising model on a $500 \times 500$ square 
lattice and quenched from a starting random spin 
configuration to zero temperature. (b) The plot of the 
scaling function $f(x)$ with $x$ in a logarithmic plot. the 
straight line has a slope 0.44 and is a guide to the eye.}{}
\label{corr}
\end{figure}

Fig.~\ref{pers} shows the decay of $P(t)$ with time in the
zero temperature quench of a Glauber Ising model on a 
$500\times500$ square lattice. Here, $P(t)$ is the fractional 
number of persistent spins. The exponent $\theta \simeq 0.22$ 
for two-dimensional Ising model \cite{stauffer}.  
Fig.~\ref{phased} shows the 
spatial distribution of the persistent spins in the same 
system at different Monte Carlo times as the  
system coarsens after it is quenched from a random initial
spin configuration to zero temperature. We can see that the 
persistent regions form a non-trivial structure which
develops with time as the system coarsens \cite{jain1}. 
The spatial correlation among the persistent sites can be 
quantified by the two point correlator $C(r,t)$ defined as 
the probability that site $(x+r)$ is persistent, given that the 
site $x$ is persistent (averaged over $x$). If $\rho(x,t)$ is 
the density of persistent sites, i.e, $\rho(x,t)=1$ if site 
$x$ is persistent at time $t$ and $0$ otherwise, then 
$C(r,t) = <\rho(x,t)>^{-1}<\rho(x,t)\rho(x+r,t)>$. Here $<...>$ 
denotes the average over different $x$ and $<\rho(x,t)>=P(t)$.  
The variation of $C(r,t)$ with $r$ corresponding to the above 
four persistent spin 
configurations is shown in fig.~\ref{corr}a. $C(r,t)$ decays 
algebraically (at late times)
with distance $r$: $C(r,t) \sim r^{-\alpha}$ up to a cutoff
length $L(t)$ which depends on time. Beyond $L(t)$, C(r,t)
is flat and independent of $r$, suggesting
that the persistent spins beyond this distance are not, on
the average, correlated. In this
region, $C(r,t) = P(t) \sim t^{-\theta} = t^{-0.22}$. At longer
times, the power law region extends to longer range and
$L(t)$, the associated correlation length for persistence
increases with time.  Consistency demands
$L^{-\alpha}(t) \sim t^{-\theta}$ implying a
power-law divergence of $L(t)$ as $L(t) \sim t^z$ where
$z = \theta/\alpha$.

The behavior of $C(r,t)$ can be summarized in the following
dynamical scaling form \cite{manoj1}   

\begin{equation}
C(r,t) = t^{-\theta}f(\frac{r}{t^z})
\label{correq}
\end{equation}   

\noindent with the scaling function $f(x) \sim x^{-\alpha}$ for 
$x << 1$ and $f(x) \simeq 1$ for 
$x >> 1$. Fig.~\ref{corr}b shows the data collapse with $z=1/2$ 
and shows the stipulated behavior of $f(x) = t^{\theta} C(r,t)$ 
with $x = r/t^z$. The value of $\alpha = 0.44$ is in accordance 
with the scaling relation: $\alpha = \theta/z$.

The power-law decay of $C(r,t)$ with $r$ implies that the
underlying structure is a scale-invariant fractal with fractal
dimension $d_f = d - \alpha$ in a $d$-dimensional system. The
persistent sites form fractal up to a length scale which
increases with time as $t^z$ and the overall spatio-temporal
evolution of the persistent regions is governed by the
above-mentioned dynamical scaling law (Eq.\ref{correq}). 
The scale-invariant structure and the power-law decay of 
persistence are related to each other and the 
fractal dimension $d_f$ provides a direct information of 
the persistence exponent $\theta$. The dynamical scaling 
seems to hold for various systems and in different dimensions 
as long as persistence shows power-law decay in time.  $d_f$ 
has indeed been determined \cite{manoj2} for various systems 
and the value of $\theta$ obtained using the scaling relation 
has been compared with that known in the literature
(see Table.~\ref{table}).

\begin{table}[tbp]
\centerline{
\hbox  {
        \vspace*{0.5cm}
        \epsfxsize=9.0cm
        \epsfbox{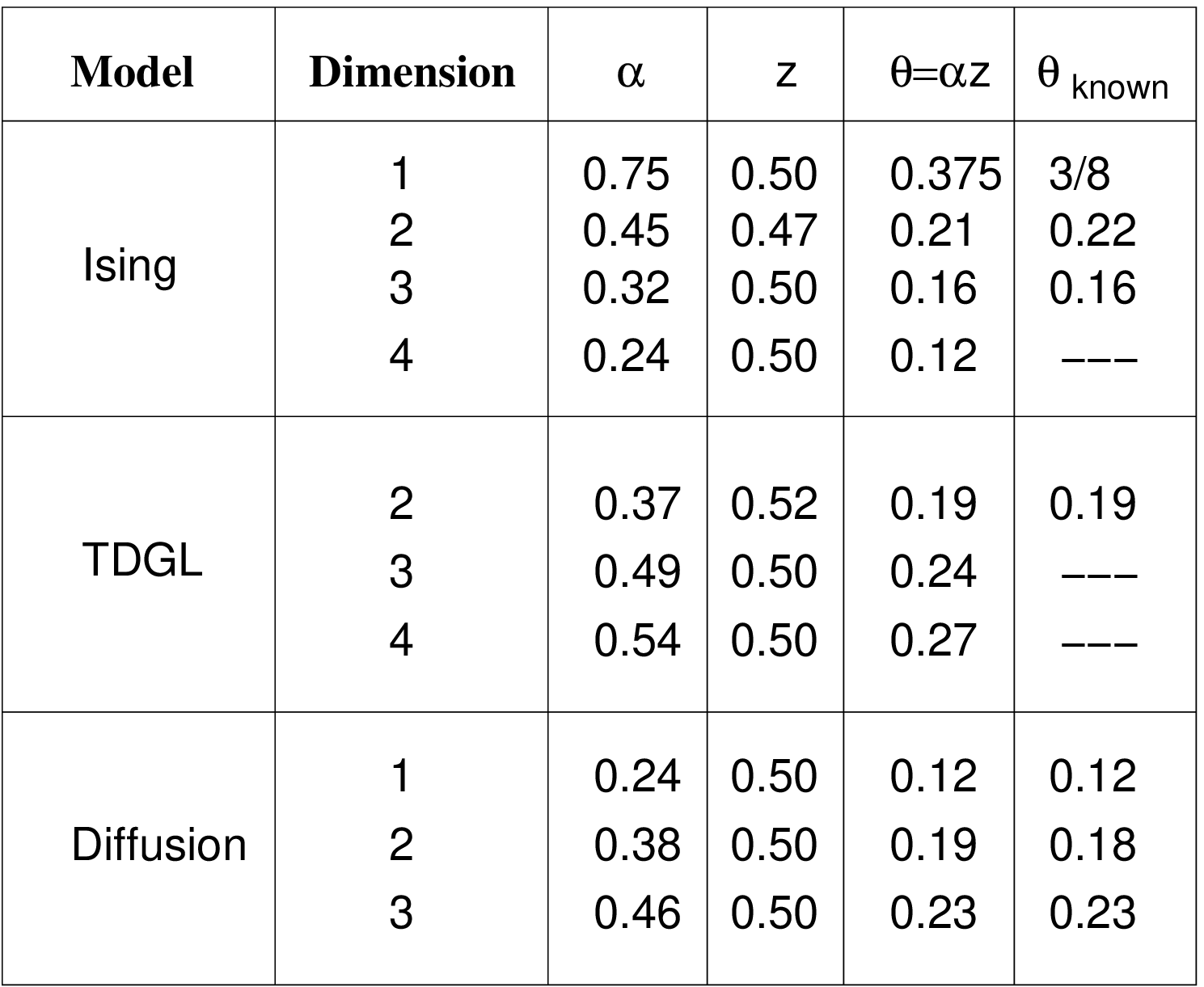}
        \vspace*{0.5cm}
       }
          }
\caption{ $\alpha$ and $z$ values shown are calculated from 
the two-point correlator and the persistence probability. $\theta$ 
obtained from the scaling relation is compared with the values of 
$\theta$ known in the literature.}{}
\label{table}   
\end{table}

Much of the dynamical scaling can be understood by looking into 
the motion of the domain walls in simple cases like 
Ising model in one dimension. As we have seen before, 
the domain walls in this case can be thought of as
Brownian particles 
$A$. The coarsening process involves the 
random motion and annihilation of
two $A$ particles ($A+A \rightarrow \emptyset$) when they come on
top of each other. Persistence
in this scenario is given by  the fraction of lattice sites that are
not visited by the particles $A$ till a certain time $t$. The decay
of persistence is the process of irreversible coalescence
of the empty intervals (segments of spin chain which are
not persistent). The distribution $n(k,t)$ of intervals of size
$k$ at time $t$ is studied using 'independent interval
approximation' (IIA). In this approximation, lengths of adjacent 
intervals are considered as uncorrelated random variables. The rate 
equation for the coalescence of the intervals can explicitly be shown 
to sustain a scaling 
solution for the empty interval distribution (see \cite{manoj3} 
for details):

\begin{equation}
n(k,t) = s(t)^{-2}\psi(\frac{k}{s(t)}) 
\label{empty}
\end{equation}

\noindent once we note that the probability of depletion of empty 
interval of size $'m'$ is zero for $m < L_D(t)$, where 
$L_D(t) \sim \sqrt{t}$ is the diffusive scale and is constant 
(time dependent) for $m > L_D(t)$ (IIA). 
The dynamical scaling ansatz (Eq.\ref{empty}) is invoked 
considering the mean empty interval length $s(t)$ proportional to 
the mean walker separation (or spin correlation 
in Ising model) $L_D(t)$: $s(t) \sim t^{1/2}$.   
Here, $\psi(x)$ is the scaling function which behaves like:  
$\psi(x) \sim x^{-\tau}$ for $x << 1$ and decays exponentially
for larger $x$. Note that $\int n(k,t)dk=P(t)$, and this  
gives the scaling relation $\tau = 2 - 2\theta$.
Eq.\ref{empty} implies that for $k<<s(t)$, 
$n(k,t) \sim t^{-\theta}k^{-\tau}$. It is the depletion of 
these intervals in this range of $k$ which predominantly 
determines the decay 
of persistence. The distribution shows a power-law decay in 
$k$ which gives rise to the fractal structure in persistence. 
The fractal dimension can be calculated if we note that 
$\tau = 2-\theta/z$ with $z=1/2$. This formulation indicates
that the diffusive scale is the scale $L(t)$ up to which the 
persistent sites show fractal structure.   
The IIA turns out to be a good approximation in the present model.
Numerical simulation results support the scaling form 
(Eq.\ref{empty}) very well with $\theta = 3/8$ and $\tau = 5/4$ 
\cite{manoj3}. The dynamical scaling 
(Eq.\ref{correq})  for the two-point correlations 
is recovered from the empty interval distribution (Eq.\ref{empty})
by Laplace transformation.

There are two relevant length scales for persistence in coaresening 
systems: the diffusive scale $L_D(t)$ and persistence scale 
$L_p(t) \sim t^{\theta}$ which is the inverse of the persistent 
fraction and gives the typical distance between two neighboring 
persistent sites. The asymptotic persistence behavior is dominated 
by the larger of the two scales. The dynamical scaling 
(Eq.\ref{empty}) is satisfied as long as $L_p(t) < L_D(t)$. 
Simulation results indicate that the mean empty interval 
length $s(t)$ depends
on the initial density $n_0$ of the $A$-particles
in the following way \cite{manoj3}:  
$$s_{n_0}(t) \sim at^{1/2} + b(n_0)t^{\theta},$$
where, $s_{n_0}(t)$ represents $s(t)$ with initial density 
$n_0$ of the $A$ particles. $b$ is found to be $<, =$ or $>0$ for 
$n_0 <, =$ or $> 1/2$. The two terms in the above expression 
plausibly originate from the two length scales 
$L_D(t)$ and $L_p(t)$. For large $n_0$, the persistent sites are 
depleted much faster and the typical interval size $s(t)$ is 
determined by the persistence scale $L_p$ only. At
late times, however, the particle density falls as a result
of annihilation reaction, the situation becomes same as that
of starting with low $n_0$ and the decisive scale crosses
over to $L_D$.

\begin{figure}[tbp]
\begin{center}
\includegraphics[width=9cm]{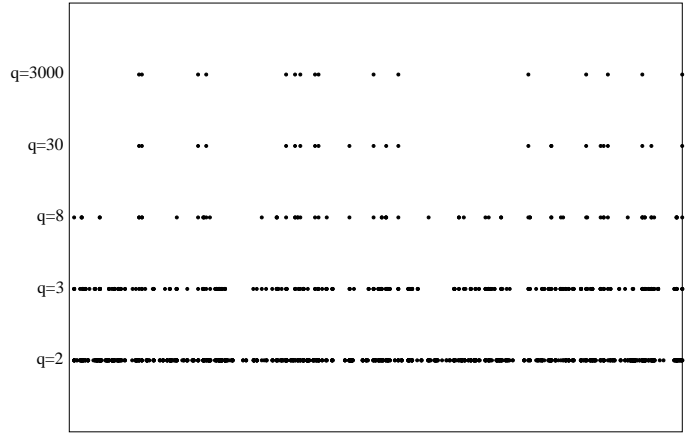}
\caption{ The persistent spins after 500 Monte Carlo steps are
shown in a one dimensional Potts model of size 20000 for different
Potts state $q$. The persistent spins at large $q$ are randomly
and sparsely distributed in space.}
\label{pottsq}   
\end{center}
\end{figure}

The interplay of these two length scales becomes apparent in the 
study of Potts model in one dimension. The domain wall
motion in Potts model depends on the Potts state
$q$: domain walls annihilate $(A+A \rightarrow \emptyset)$
with probability $1/(q-1)$ and coalesce $(A+A \rightarrow 1)$
with probability $(q-2)/(q-1)$. For $q=2$, these probabilities
correspond to those in Ising model. For larger $q$, the
coalescence reaction is more probable and particles $A$
decay slowly. The persistence probability (the fraction
of sites that were never crossed by $A$) decays faster, as a result, 
but retains its algebraic form: $P(t) \sim t^{-\theta(q)}$, where
$\theta(q)$ is now $q-$dependent and increases with $q$ 
(see Eq.\ref{potts}). 
The value of $q$ corresponding to $\theta =1/2$ is 2.70528...,
so $\theta(q) > 1/2$ for any $q \ge 3$ and $L_p$ should be
the dominant length scale in that case. The characteristic distance of
separation $<s(t)>$ between clusters for persistent sites
is then given by $<s> =$ max$[L_D,L_p]$ \cite{bray1,manojsolo}. 
Fig.~\ref{pottsq} shows the 
persistent spins in a one-dimensional Potts model for various 
$q$ values. For large $q$, the persistent spins are sparsely 
distributed in space and their average number decreases with the 
increase of the system size ($d_f=d-\theta/z$ is negative, $z$ 
remains $1/2$).

\begin{figure}[tbp]
\vspace*{-1.0in}
\centerline{
\hbox  {
        \vspace*{0.5cm}
        \epsfxsize=9.0cm
        \epsfbox{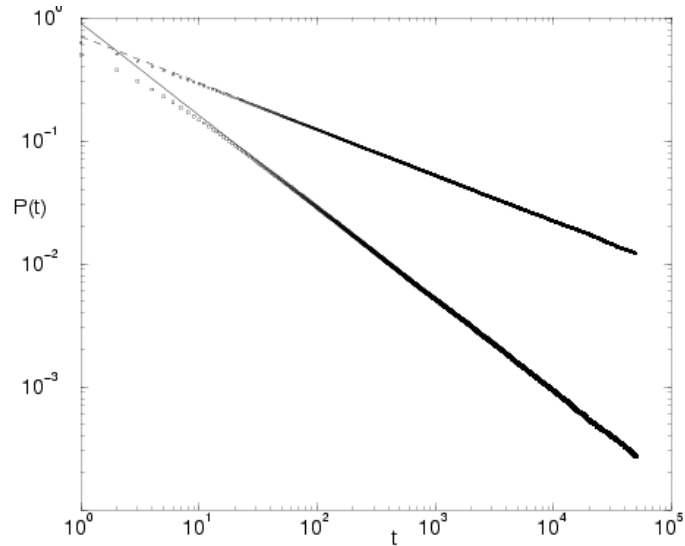}
        \vspace*{0.5cm}
       }
          }
\caption{The persistence probability $P(t)$ in a one-dimensional
Ising model is plotted against time $t$ in a logarithmic scale.
The lower curve is for synchronous spin updating and the upper
one is for asynchronous updating scheme.  The solid and dashed
lines fitted to these cures are guide to the eye and have slopes
0.75 and 0.375 respectively.}{}
\label{shukla}   
\end{figure}

Next, we address, how persistence exponent depends on the dynamics 
of the system. Persistence is related to the dynamical evolution 
of a system, Hence
the persistence exponent $\theta$ may show dependence on the
detailed microscopic updating rule in system just as
the dynamical exponent $z$ in critical phenomena depends
on the updating rule imposed on the system. We discuss
below how two most commonly applied updating rules, namely
synchronous and asynchronous spin updating can alter the
exponent $\theta$ and how one can understand the effect of 
updating scheme on persistence in the simple case of zero temperature
quench of a Potts chain. In this case, persistence decay remains
algebraic for both synchronous and asynchronous
spin updating, but the persistence exponent $\theta_s(q)$ in the
case of synchronous dynamics is exactly twice of
$\theta_a(q)$, the persistence exponent for asynchronous
spin updating scheme \cite{shukla,menon}. This result is
valid for all $q$. For example, for $q=2$ or equivalently in
Ising model, $\theta_a(2) = \frac{3}{8}$ \cite{derrida2} and 
$\theta_s(2)=3/4$ \cite{shukla} (see fig.~\ref{shukla}).

Inspection of the spins at different times under synchronous
updating scheme shows the formation of large regions where the
spins are arranged as $10101010..$ where 1 and 0 are the two
states of an Ising spin in Ising model. All the spins in such
regions are unstable and flip at every time step. As a result, the 
persistence
decay is much faster. These unstable regions are formed because 
of the unbinding of the two zero-field spins that constitute
a domain wall. In asynchronous updating rule, such unstable
regions cannot be formed. For example, in a spin configuration 
like $111000$, the third and the fourth spins at the junction 
of 0 and 1 spin domains have zero local field. In asynchronous 
dynamics, these zero-field spins 
remain always bound together and form a domain wall. During 
coarsening, these domain walls move and are represented by 
particles $A$ in corresponding reaction diffusion system as has 
been mentioned before. In synchronous dynamics, the zero field 
spins across a domain wall can move apart with the 
formation of unstable spin 
cluster in between them, like in 111010101000. Here, the third 
and the tenth spins have zero local field and these spins 
individually, rather than the domain walls, act as the reacting 
particles $A$, as far as persistence is concerned, in the 
corresponding reaction diffusion system. A 
zero-field site should be recognized as a site which can take any 
one of the neighboring spin states with probability $1/2$ at any 
time step. In Potts model with $q>2$, the second unstable spin 
in a spin arrangement like 123 (1,2,3... are spins with 
$q$=1, 2 and 3) is also a zero-field spin and 
hence, a reacting particle $A$, to complete the analogy with the 
reaction diffusion process. It is easy to see, that, starting from 
an initial spin configuration, a spin becomes (with probability 
$1/2$ for the central spin in a spin configuration like $100$ 
and with probability 1 in a configuration like $123$) non-persistent 
only when it becomes a zero-field spin in course of evolution.

\begin{figure}[tbp]
\vspace*{-2.5in}
\centerline{
\hbox  {
       \vspace*{0.5cm}
         \epsfxsize=12.0cm
        \epsfbox{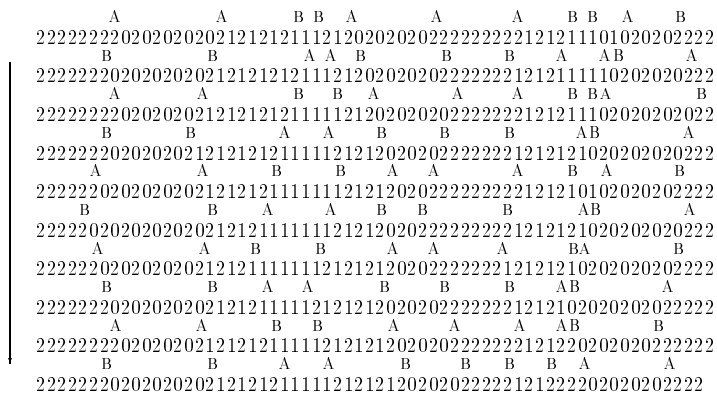}
        \vspace*{-1.5cm}
       }
          }
\caption{Evolution of a typical spin configuration in 3-states
Potts model under synchronous updating scheme is shown. The
configurations are separated by one time step ; earlier
times appear at the top row. The spin states are shows as 1,2
and 3. The reactant particles are shown as $A$ or $B$ according
to the sublattice on which they are present at any particular time.
The two-sublattice decoupling is an exact feature of the model in
which the reactant particles on each sublattice, and at
alternating times, are identified with the diffusing and
annihilating random walkers.}{}
\label{gautam}   
\end{figure}

Evolution of a spin configuration in 
a 3-state Potts model is illustrated in fig.~\ref{gautam} where 
the reacting particles are labelled as $A$ or $B$ according to 
the sublattice on which they are present at a time $t$. Note 
that a particle $A$ at time $t$ becomes the particle $B$ at time 
$(t+1)$. 
Considering a spin configuration like $123122$, where two zero-field 
sites are side-by-side at positions 3 and 4, it is easy to see 
that these two sites can never react. The dynamics may make a 
$AB$ pair a $BA$ pair, in which case we may say that the 
particles have tunnelled through. Whereas, two zero-field 
sites at two alternating sites (two $A$ or $B$ sites) 
like the sites at position 3 and 5 in the configuration 
$3331333$ can react ($3331333 \rightarrow 3333333$).
A careful study shows that the dynamics of the zero-field sites 
under synchronous dynamics can be mapped \cite{menon} exactly 
to the reaction-diffusion dynamics of the particles $A$ or $B$ on one or the 
other sublattices at times $t=1, 3, 5 ...$ or at $t=2, 4, 6, ...$.
The two reaction-diffusion processes    
are completely independent of each other and the 
dynamics of $A$ (or $B$) alone is analogous to the dynamics  
of domain walls in asynchronous spin updating scheme. The system, 
as a result, can be looked upon as comprising of two decoupled 
reaction-diffusion systems. For random initial starting configurations, 
there is no
correlations in the initial placement of $A$ and $B$ particles, 
the joint probability that a given site is persistent with respect 
to the motion of both $A$ and $B$ particles simply factors into 
the product of independent probabilities at late times:
$$P_{syn}(T) \sim \frac{1}{t^{\theta_s}} = P_{asyn}P_{asyn} \sim \frac{1}{t^{\theta_a}}\frac{1}{t^{\theta_a}}$$ 
giving the result 

\begin{equation}
\theta_s = 2 \theta_a.  
\label{syn}
\end{equation}

\begin{figure}[tbp]
\vspace*{-2.0in}
\centerline{
\hbox  {
        \vspace*{0.5cm}
        \epsfxsize=9.0cm
        \epsfbox{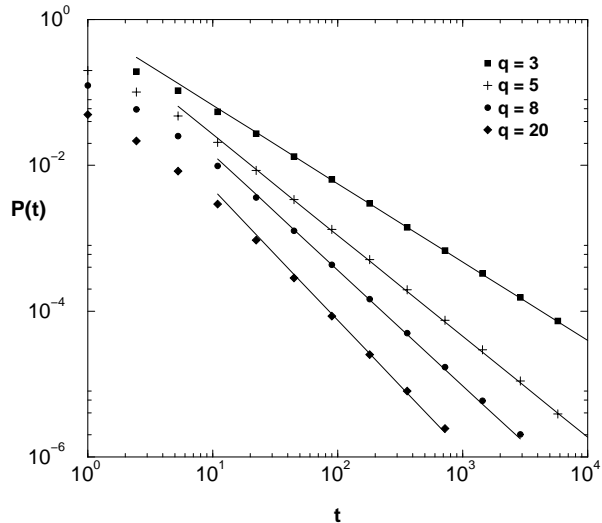}
        \vspace*{0.5cm}
       }
          }
\caption{The persistence probability $P(t)$ in a one-dimensional
$q-$state Potts model evolving from random initial configurations
under synchronous dynamics is plotted against time $t$ in a
logarithmic scale for $q = 3, 5, 8$ and $20$. The 
lines fitted to these cures are guide to the eye and have slopes
$\theta_p(q=3) \simeq 1.076$, $\theta_p(q=5) \simeq 1.386$,
$\theta_p(q=8) \simeq 1.588$ and $\theta_p(q=20) \simeq 1.821$.}{}
\label{gautam1}   
\end{figure}

\begin{figure}[tbp]
\vspace*{-1.0in}
\centerline{
\hbox  {
        \vspace*{0.5cm}
        \epsfxsize=9.0cm
        \epsfbox{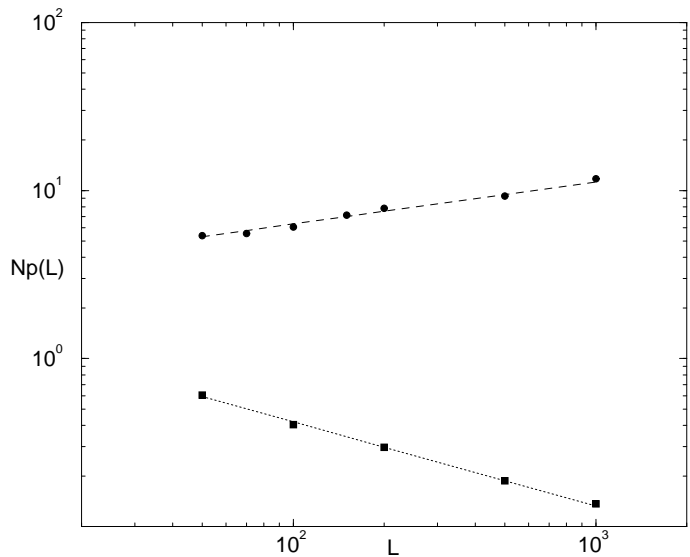}
        \vspace*{0.5cm}
       }
          }
\caption{The total number of persistent sites 
$N_p(L)=LP(L,t\rightarrow\infty)$ left in a one-dimensional 
Ising system at late times is plotted against $L$ on a 
logarithmic scale for both synchronous (circles) and 
asynchronous (squares) dynamics. The straight lines fitted 
to these points have slopes $-0.5$ and $0.25$ respectively.}{}
\label{gautam2}   
\end{figure}

Fig.~\ref{gautam1} shows the simulation results for the 
power-law decay of persistence probabilities in a 
one-dimensional Potts model for Potts states $q = 3, 5 ,8$ 
and $20$ under synchronous spin updating rule. Persistence 
exponent $\theta_s(q)$ is found to be twice of $\theta_a(q)$ 
which is obtained from the exact expression Eq.\ref{potts}. 
For Ising model $\theta_a(2)=3/8$, so $\theta_s(2) = 3/4$. 
The dynamical scaling (Eq.\ref{correq}) still holds good 
\cite{shukla} with $z=1/2$ for synchronous spin updating of 
the spins. The  fractal dimension of the persistent spin 
regions becomes $d_f = d -z\theta = -0.5$. The negative 
fractal dimension is attributed to the fact that we are 
looking at the persistence of a site under two independent 
processes: each of which forms a fractal with fractal 
dimension $d_f=0.25$ (with $z=2$ and $\theta=3/8$). The 
intersection of these two fractals represents those sites 
which are persistent with respect to the motion of both 
$A$ and $B$ particles. The dimension of the intersection 
set is then $2d_f-d = -0.5$, as above. One implication of 
negative fractal dimension is that both the average number 
and average density of persistent sites in a system of size $L$
decay with $L$: a large system has fewer persistent sites at 
sufficiently long times. Fig.~\ref{gautam2} shows the 
simulation result of the total number of persistent sites 
in one-dimensional Ising model for both 
synchronous and asynchronous spin dynamics. For asynchronous 
spin dynamics, average number of persistent sites increases with 
the system size while its density decreases.

Since persistence probes the detailed time evolution of the 
system it is hard to expect any universal behavior in persistence.
In fact, in many cases, $\theta$ is found to depend on the lattice
coordination number or the precise choice of the
inter-particle interaction \cite{hinrichsen,jain}.
We have seen how persistence exponent $\theta$ depends on the
dynamics or the underlying updating rule in the system. 
In this background, it is intriguing that for a varieties of 
$1+1$-dimensional directed percolation processes,   
$\theta$ is observed to be $\sim 1.5$ \cite{hinrichsen}. 
Directed site and bond percolation \cite{grass}, cellular
automata such as the Domany-Kinzel model \cite{domany} and
Ziff-Gulari-Barshad model \cite{ziff}, contact processes
\cite{liggett}, coupled map lattices showing spatio-temporal
intermittency \cite{pomeau} are some of the examples
of directed percolation processes. These systems show, on changing 
a relevant parameter, an absorbing state phase transition. 
Starting from a random initial configuration, the
dynamics leads, in some parameter space, to an absorbing state 
(where no further evolution of the system can take place). On 
changing the parameter, one finds a continuous transition to a 
phase where activity never ceases. The critical parameter value 
corresponds to a phase transition point. Absorbing state phase 
transitions, in general, belong to the class of directed percolation 
(provided the absorbing phase is unique, the processes short-ranged, 
there is no additional symmetries or randomness and the order 
parameter is positive and has single component \cite{grass}).

\begin{figure}[tbp]
\vspace*{-1.0in}
\centerline{
\hbox  {
        \vspace*{0.5cm}
        \epsfxsize=6.0cm
        \epsfbox{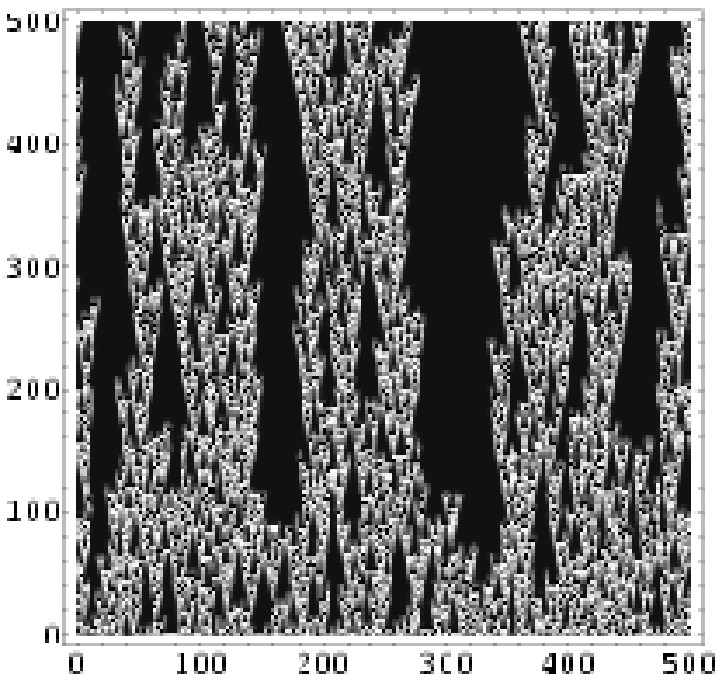}
        \hspace*{1.2cm}
        \epsfxsize=6.0cm
        \epsfbox{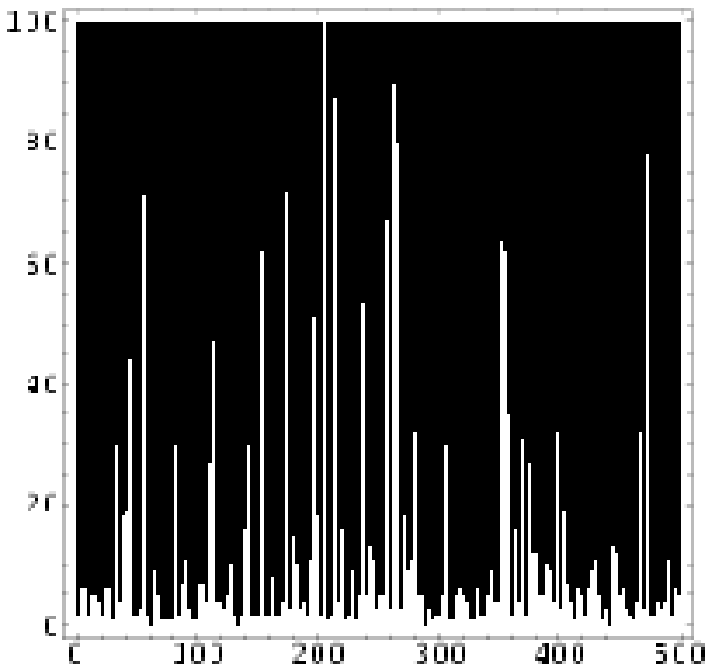}
        \vspace*{0.5cm}
       }
          }
\centerline{
\hbox  {
        \vspace*{0.5cm}
        \hspace*{1.0cm}  
        \epsfxsize=0.5cm
        \epsfbox{fig1a.eps}
        \hspace*{6.8cm}
        \epsfxsize=0.5cm
        \epsfbox{fig4b.eps}
        \vspace*{0.5cm}
       }
          }

\caption{Time evolution (given by the y-axis) of the one-dimensional
coupled map of size $L=500$ at the critical point. (a) The density
plot of the actual $x_{i,t}$ values (the absorbing regions appear
dark). (b) Corresponding picture of the persistent regions 
evolving in time.}{}
\label{sudeshna1}   
\end{figure}

As an example of absorbing state phase transition of directed
percolation universality class consider a one-dimensional coupled
map lattice, with on-site circle maps coupled diffusively to nearest
neighbors \cite{sinha}:

\begin{equation}
x_{i,t+1} = f(x_{i,t}) + \frac{\epsilon}{2}
(f(x_{i-1,t})+f(x_{i+1,t}) - 2 f(x_{i,t})) \ \ \ {\rm
mod} \ \ 1  
\label{coupled}  
\end{equation}

\noindent where $t$ is the discrete time index, and $i$ is the site index.
The parameter $\epsilon$ measures the strength of the diffusive
coupling between site $i$ and its two neighbors. The on-site map is 
$$f(x) = x + \omega - \frac{k}{2\pi} sin(2\pi x).$$
The fixed point solution
$$x^{*} = \frac{1}{2\pi}sin^{-1}(\frac{2\pi \omega}{k})$$ 
corresponds to the unique absorbing state. Active sites have local
value of $x$ different from $x^*$. For appropriate choice of the
parameters $k, \omega$ and $\epsilon$ one obtains a critical state.
Otherwise the dynamics leads to a completely inactive
laminar state where not a single site evolves any further or
to a turbulent state leading to an active phase without any 
spatio-temporal regular structure of the active sites. Critical 
point marks the onset of spatio-temporal intermittency 
(see fig.~\ref{sudeshna1}).

\begin{figure}[tbp]
\vspace*{-1.5in}
\centerline{
\hbox  {
        \vspace*{0.5cm}
        \epsfxsize=9.0cm
        \epsfbox{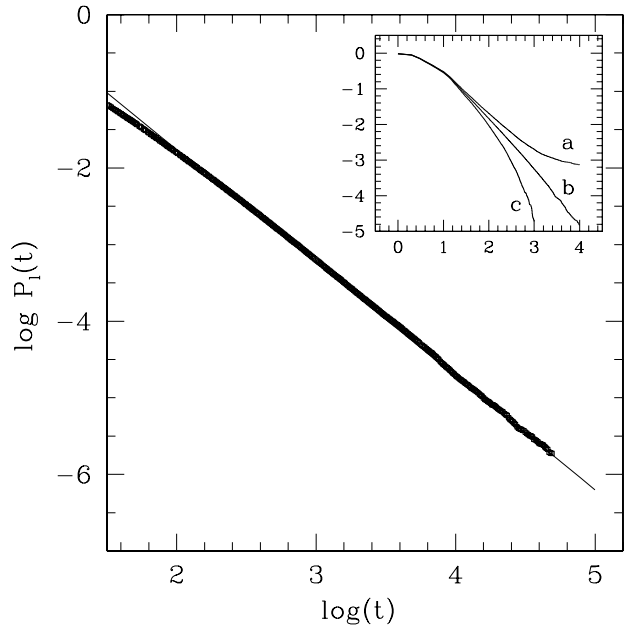}
        \vspace*{0.5cm}
       }
          }
\caption{Persistence probability $P(t)$ vs time $t$ in
logarithmic scale at the critical point. The straight line fitted
to the curve has a slope 1.49. Inset shows the log-log
plot of $P(t)$ below, at and above the critical point. }{}
\label{sudeshna2}   
\end{figure}

The critical properties of the coupled map lattice at the spatio-temporal 
intermittency are found numerically to be in the universality class 
of directed percolation \cite{neelima}. Persistence in coupled map lattices 
is defined in terms of the probability that a local site variable 
$x_{i,t}$ does not cross the fixed point value $x^*$ up to time 
$t$. Fig.~\ref{sudeshna1}(b) shows the evolution of the persistent 
sites with time. The persistent probability decays with time as 
$P(t) \sim t^{-\theta}$ (see fig.~\ref{sudeshna2}) with 
$\theta = 1.49 \pm 0.02$.  This value of $\theta$ is very close 
to the value $\sim 1.5$ obtained in Domany-Kinzel model 
\cite{hinrichsen}. In Ziff-Gulari-Barshad model and contact 
processes $\theta$ seems to be $\sim1.5$ both in dimensions one and 
two \cite{albano}. For dimensions greater that two, $\theta$ 
decreases with dimension. Based on these observations, it is conjectured 
that $\theta=3/2$ for directed percolation processes \cite{hinrichsen}. 
In $1+1$-dimensional directed bond percolation process with 
an absorbing boundary \cite{essam}, it has been 
shown that the persistence probability is exactly equal  
to the return probability of the process with an 
active source to return to a state where all sites except the 
source are inactive \cite{hinrichsen}. Though the proof is done 
only for directed bond percolation process, simulation results indicate 
the validity of the mapping for various transition points in 
the Domany-Kinzel model. The seeming universal persistence behavior 
for directed percolation processes should be studied further.


We have discussed the spatial correlation that arises among 
the persistent sites in interacting many-body systems evolving 
in time. This spatial correlation gives rise to fractal structures 
in persistent regions, dynamical scaling and power-law decay 
of persistence. The persistence behavior depends crucially on the 
dynamics of the system. In one-dimensional Potts model, we have 
discussed how the persistence decay depends on the Monte Carlo 
updating rules and the exact relation between the persistence 
exponent in the case of asynchronous spin updating rule to 
that in synchronous spin updating rule. Finally, we have 
discussed the universality that is observed in the persistence 
behavior in several systems at the directed percolation 
transition. The correlation among persistent regions should be 
studied in more details for directed percolation processes
to get ideas about the possible origin of the observed 
universal behavior of persistence.    

\vspace{0.5cm}  

\noindent Acknowledgements: I am grateful to 
Prof. D. Dhar, Prof P. Shukla, Prof. S. Sinha and Dr. G. Manoj
for their valuable suggestions and comments. I thank Dr. R. Roy 
for critically reading the manuscript.

\begin{flushleft} {\bf References:} \end{flushleft} 

\end{document}